\documentclass[useAMS,usenatbib]{mnras}

\usepackage{graphicx}
\usepackage{amsmath,amssymb}
\usepackage{lmodern}
\usepackage{comment}
\usepackage{float}
\usepackage{xcolor}

\title[ALPs and the VHE Spectrum of M87]{Axion-like Particles and their Possible Impact on the Very High-Energy Spectrum of M87 Observed by LHAASO}

\author[A. Pratts et al.]{
A. Pratts,$^{1}$\thanks{Contact~e-mails: \href{mailto:yoba\_m\_t\_a@ciencias.unam.mx}{yoba\_m\_t\_a@ciencias.unam.mx}~(AP); \href{mailto:daniel\_avila5@ciencias.unam.mx}{daniel\_avila5@ciencias.unam.mx}~(DAR)}
D. Avila Rojas,$^{2\star}$
J. Serna-Franco,$^{1}$
M.M. González,$^{2}$
R. Alfaro,$^{1}$\\
$^{1}$Instituto de F\'{i}sica, Universidad Nacional Autónoma de México, Ciudad de México, México\\
$^{2}$Instituto de Astronom\'{i}a, Universidad Nacional Autónoma de México, Ciudad de México, México
}

\pubyear{2025}

\begin{document}
\label{firstpage}
\pagerange{\pageref{firstpage}--\pageref{lastpage}}
\maketitle

\begin{abstract}
The detection of very high-energy (VHE) gamma rays from the active galaxy M87 by LHAASO, showing a possible spectral hardening around $20$ TeV, motivates the search for new physics beyond standard emission models. One promising candidate is axion-like particles (ALPs), hypothetical pseudo-scalar bosons that can oscillate into photons in the presence of cosmic magnetic fields. In this work, we investigate whether photon-ALP oscillations and an additional ALP-induced component can account for the tentative hardening observed in M87’s VHE spectrum. We model the propagation of photons and ALPs through the jet, the Virgo cluster, the intergalactic medium, and the Galactic magnetic field, over a broad ALPs parameter space. Our statistical analysis finds that, with current LHAASO data, the inclusion of an ALPs component yields only a modest improvement over a standard scenario (maximum significance $\sim$1.56$\sigma$). However, if future observations transform current flux upper limits at tens of TeV into measured fluxes, the significance could reach $\sim$3$\sigma$, providing potential evidence for ALP-induced effects. Our results suggest that M87 remains a promising target to test fundamental physics, and upcoming VHE data could play a key role in probing ALPs parameter space.
\end{abstract}


\section{Introduction}
The development of more advanced TeV observatories has enabled numerous studies, thereby increasing our understanding of the various processes associated with the most energetic events in the Universe, such as supernovae and active galactic nuclei (AGNs). Precise data on very high-energy emissions have provided a valuable tool not only for studying these sources themselves but also for testing theories about the fundamental composition of the Universe. One prominent example is axion-like particles (ALPs), hypothetical particles proposed initially within the context of quantum chromodynamics. A distinctive characteristic of these low-mass particles ($<$ eV) is their ability to oscillate into photons in the presence of magnetic fields. These oscillations depend, among other factors, on the energy, whether it is a photon or an ALP. The study of sources emitting photons at very high energies offers a unique opportunity to test the potential impact on the source spectra observed from Earth of ALPs coupled to photons. 
One of the most extensively studied sources is the Active Galactic Nucleus (AGN) M87, a radio galaxy located at the center of the Virgo Cluster, with a redshift of z = 0.0044. It represents an interesting object of study for extragalactic astrophysics, particularly in the investigation of dark matter, because analysis of its stellar and gas dynamics reveals a surprisingly high mass-to-light ratio ($M/L$), indicating that a significant fraction of its total mass is composed of dark matter \citep{2011ApJ...729..119G,2019PhRvD.100d4012J}. In the inner region, $M/L$ ratio rises up to $8.65\pm 0.38$, suggesting that dark matter constitutes almost $\sim(67\pm2)\%$ \citep{2023ApJ...945L..35L}.

The distribution of this dark matter within M87 is a key factor in understanding its nature and properties. Cosmological simulations and theoretical models predict that dark matter tends to concentrate in the centers of galaxies, forming dense halos that follow characteristic density profiles, such as the Navarro-Frenk-White (NFW) profile \citep{1996ApJ...462..563N}. 
In such an environment of high density and strong magnetic fields, 
ALPs can convert into photons and vice versa, opening the possibility of detecting their presence through signatures in high-energy gamma-ray observations \citep{2014JCAP...09..003M, 1988PhRvD..37.1237R, 2011PhRvD..84j5030D, 2009JCAP...12..004M, 2021PhRvD.103d3018B}.

Gamma-ray observatories have conducted multiple monitoring campaigns, detecting emission from quiescent and flare states from M87 \citep{2006Sci...314.1424A, 2008ApJ...679..397A, 2008ApJ...685L..23A, 2010ApJ...716..819A, 2012ApJ...746..141A}. Moreover, the LHAASO Collaboration recently published the detection of VHE gamma-ray emission from M87, with a spectrum extending up to 20 TeV and reported a possible hardening at $\sim20$ TeV \citep{2024ApJ...975L..44C}. This result was used to constrain the intensity of the extragalactic background light (EBL), presenting the intrinsic spectrum of M87 assuming the \citet{2017A&A...603A..34F} model. The spectral hardening observed by LHAASO is not statistically significant with its currently available data; nevertheless, the HAWC Observatory has also published the measured spectrum of M87, which also shows a possible hardening at $\sim20$ TeV \citep{2025arXiv250616031A}. Both observations, from LHAASO and HAWC, motivate the study of potential hardening as an additional component under an ALPs scenario.

In this work, we explore the hypothesis that an extra spectral component from ALPs oscillations contributes to the VHE gamma-ray emission observed from M87. By assuming the existence of this component, we seek to account for the hardening observed at $\sim20$ TeV, 
and to establish the spectral features of this ALPs component. 

\section{Axion-like Particles and Photon Mixing}

Axion-like particles (ALPs) are pseudo-scalar bosons that arise in various extensions of the Standard Model, notably in string theories and compactified extra-dimensional models. Unlike the quantum chromodynamics (QCD) axion, ALPs are not constrained to solve the strong CP problem and hence may have a broad range of masses $m_a$ and coupling constants $g_{a\gamma}$ \citep{1983PhLB..120..133A,2006JHEP...05..078C,2010PhRvD..81l3530A}.

The interaction Lagrangian for the photon-ALP system is given by:

\begin{equation}
\mathcal{L}_{\rm int} = -\frac{1}{4}g_{a\gamma} a F_{\mu\nu}\tilde{F}^{\mu\nu} = g_{a\gamma} a \mathbf{E}\cdot\mathbf{B},
\end{equation}

where $a$ is the ALP field, $F_{\mu\nu}$ is the electromagnetic field tensor, $\tilde{F}^{\mu\nu}$ its dual, and $g_{a\gamma}$ is the ALP-photon coupling constant with units of the inverse of the energy \citep{PhysRevD.37.1237}.

The evolution of a photon/ALP beam propagating in a magnetic field is governed by the equation of motion:

\begin{equation}
\left(i\frac{d}{dz} + E + \mathcal{M}(z)\right)\Psi(z) = 0,
\end{equation}

where $\Psi(z) = (A_x, A_y, a)^T$ is the propagation state and $\mathcal{M}(z)$ is the mixing matrix given by:

\begin{equation}
\mathcal{M} =
\begin{pmatrix}
\Delta_{\gamma} & 0 & \Delta_{a\gamma,x} \\
0 & \Delta_{\gamma} & \Delta_{a\gamma,y} \\
\Delta_{a\gamma,x} & \Delta_{a\gamma,y} & \Delta_a
\end{pmatrix},
\end{equation}

with

\begin{align}
\Delta_a &= -\frac{m_a^2}{2E}, \\
\Delta_{a\gamma} &= \frac{1}{2}g_{a\gamma} B_T, \\
\Delta_{\gamma} &= -\frac{\omega_{\rm pl}^2}{2E},
\end{align}

where $B_T$ is the transverse magnetic field component, and $\omega_{\rm pl}$ is the plasma frequency given as $\omega_{\rm pl} = \sqrt{4\pi\alpha n_e/m_e}$, with $n_e$ being the electron density.

In the single domain approximation, the photon survival probability is given by \citep{2009JCAP...12..004M,2011PhRvD..84j5030D,2018PhRvD..98d3018G}:

\begin{equation}
P_{\gamma\rightarrow\gamma}(E) = 1 - P_{\gamma\rightarrow a}(E) = 1 - \sin^2(2\theta)\sin^2\left(\frac{\Delta_{\rm osc} L}{2}\right),
\end{equation}

where

\begin{align}
\tan(2\theta) &= \frac{2\Delta_{a\gamma}}{\Delta_a - \Delta_{\gamma}}, \\
\Delta_{\rm osc} &= \sqrt{(\Delta_a - \Delta_{\gamma})^2 + 4\Delta_{a\gamma}^2},
\end{align}

where $P_{\gamma\rightarrow\gamma}(E)$ is the survival probability of a photon converted to an ALP, and back to a photon ($P_{\gamma\rightarrow a\rightarrow\gamma}(E) = P_{\gamma\rightarrow\gamma}(E)$), and $P_{\gamma\rightarrow a}(E)$ is the survival probability of photons converted into ALPs that could not convert back in their way from the source to the observer.

Multiple domains and stochastic magnetic fields require numerical integration across varying $B$ fields and densities.

\section{Gamma-ray Contribution from an ALPs Extra Component}
As mentioned above, photons emitted by extragalactic sources can oscillate into ALPs; such an oscillation can be triggered by their interactions with magnetic fields along their path to Earth \citep{2018JHEAp..20....1G}. Eventually, some of these ALPs can reconvert into photons. As a result, the original gamma-ray spectrum may exhibit deviations from a well-defined function (for example, from a power law) as signatures of ALPs production. This scenario has been explored in several studies.
In addition to the photon-ALP oscillations described above, other possible scenarios producing extra photons involves the direct emission of ALPs from first-order phase
transition \citep{2023PhLB..83937824N}, and from magnetized regions near the black hole's event horizon in AGNs \citep{2025PhRvD.111f3003P}. Therefore, we adopt the hypothesis that there is a population of ALPs, and we model its spectrum as a simple power-law distribution.

\begin{equation}
\phi_{a}(E) = \rm{N}_a \left(\frac{E}{E_{0,a}}\right)^{-\Gamma_a},
\end{equation}

where $\rm{N}_a$ is the normalization, $\Gamma_a$ is the spectral index, and $E_{0,a}$ is the pivot energy fixed at 3 TeV.

The underlying physical processes that accelerate or produce ALPs up to TeV energies remain outside the primary scope of this work.

Upon reaching the Milky Way, some of these ALPs may convert into gamma-ray photons, contributing to the observed spectrum. The overall observed photon spectrum becomes the sum of the attenuated primary photon spectrum and the converted ALPs signal:

\begin{equation}
\phi_{\rm total}(E) = \phi_0(E) P_{\gamma\rightarrow\gamma}(E)  + \phi_{a}(E) P_{a\rightarrow\gamma}(E),
\end{equation}\label{eq:obsflux}

where $\phi_0$ is the intrinsic spectrum of the source taken as reported by LHAASO as a simple power law with a spectral index of $2.37\pm0.14$ and a flux normalization at $3$ TeV of $(3.30\pm0.44)\times10^{-14}\;\rm{TeV}^{-1}\;\rm{cm}^{-2}\;\rm{s}^{-1}$. The attenuation of the gamma-ray flux due to the EBL is taken into account in terms of the photon survival probability and the ALP conversion probability.

We explore the values of the ALPs spectrum parameters, $\rm{N}_{a}$ and $\Gamma_a$, to find those that reproduce the spectral hardening observed by LHAASO. 
This approach enables us to test the viability of a multi-component emission model for describing the VHE spectrum of M87.


\section{Spectral Fitting and Photon-ALP Oscillations}

To evaluate the impact of ALPs on the VHE gamma-ray spectrum of M87, we perform a statistical comparison between two models: an observed resulting spectrum with and without the contribution from ALP-photon mixing. Our framework accounts for ALPs-induced spectral modifications arising from both standard photon-ALP oscillations along the line of sight and a contribution from an ALPs population as described in Section 3.

\subsection{Magnetic Environments and Survival Probabilities}
Photon-ALP oscillations are simulated using the Python-based package \texttt{gammaALPs}\footnote{\url{https://gammaalps.readthedocs.io/en/latest}} \citep{2022icrc.confE.557M}. The astrophysical magnetic environments considered in our analysis are the jet region of the source, the intracluster medium (ICM), the intergalactic medium (IGM), and the Milky Way magnetic field (GMF). The relativistic jet from M87 is modeled as a helical magnetic field with a strength comparable to that reported by \citet{2009ApJ...699..305H} for the HST-1 emission region. The ICM is modeled as a turbulent magnetic field representative of the Virgo Cluster \citep{2024icrc.confE.908C}, using a Gaussian random field with a Kolmogorov-like power spectrum and radial scaling $B(r) \propto n_e(r)^\eta$ \citep{2024icrc.confE.908C}. The IGM is modeled as a uniform magnetic field across Mpc-scale domains and a strength according to the lower bounds reported by \citet{2010ApJ...722L..39A} and \citet{2023A&A...670A.145A}. The GMF is described using the \citet{2012ApJ...757...14J} model, which includes both large-scale and turbulent components. Finally, the effects of absorption by EBL are considered using the \citet{2017A&A...603A..34F} model to be consistent with the results of LHAASO. Table \ref{tab:params} shows the values of the magnetic field strengths used for each astrophysical environment.

\begin{table}
    \caption{
    The magnetic field values considered for the photon-ALP oscillation probability correspond to those present in the astrophysical environments of the jet, ICM, IGM, and GMF.}
    \label{tab:params}
    \centering
    \begin{tabular}{lc}
    \hline\hline
        \textbf{Environment} & \textbf{Magnetic Field} \\
        \hline
        Jet magnetic field ($B_{0,\text{Jet}}$) & $0.6$ \;\text{mG} \\
        Virgo cluster ICM ($B_{0,\text{ICM}}$) & 34.2 $\mu\text{G}$ \\
        IGM ($B_{0,\text{IGM}}$) & $10^{-9}$ $\mu\text{G}$ \\
        \hline\hline
    \end{tabular}
\end{table}

For each ALP candidate with a given mass and coupling factor \((m_a, g_{a\gamma})\), we calculate the photon survival probability \( P_{\gamma \to \gamma}(E) \) and the ALP-to-photon reconversion probability \( P_{a \to \gamma}(E) \) across the line of sight. To account for statistical fluctuations, these calculations were performed for different magnetic-field realizations (seeds).

\subsection{Fitting Strategy and Data Treatment}

The total expected photon flux \( \phi_{\rm total}(E) \) is computed using equation \ref{eq:obsflux}, considering both contributions: from the source and the ALP-photon mixing. We calculate the log-likelihood of the resulting photon flux for each ALP candidate by comparing it to the observed fluxes reported by the LHAASO Collaboration. The log-likelihood function takes the general form:

\begin{equation}
\log \mathcal{L}(\theta) = -\frac{1}{2} \sum_i \left( \frac{\phi^{\rm obs}_i - \phi_{\rm model}(E_i; \theta)}{\sigma_i} \right)^2,
\end{equation}

where $\theta$ represents the spectral parameters $\rm{N}_0$, $\Gamma$, $\rm{N}_a$ and $\Gamma_a$. The survival, or reconversion, probabilities are fixed per $(m_a, g_{a\gamma}, \text{seed})$. For the upper limits in LHAASO's data, the corresponding log-likelihood terms are modified by integrating the Gaussian probability density up to the flux limit, as described in the standard treatment of censored data \citep{1985ApJ...293..192F}. The ALPs candidates that predict a photon flux in the VHE range higher than the reported upper limits are discarded.

\subsection{Parameter Scan and Statistical Comparison}\label{subsec:4.3}

We perform a full scan over a 2D grid of the ALPs parameters, with 30 logarithmic steps in both \( m_a \) and \( g_{a\gamma} \), totaling 900 candidates explored. For each candidate, 50 random magnetic field realizations are simulated, leading to 45,000 independent spectral fits. For each set of realizations, the realization that best describes the LHAASO spectrum is kept.

We optimize the spectral parameters of the ALPs candidates by maximizing \( \log \mathcal{L} \), firstly identifying a primary region by global exploration \citep{cite-key}, and then refining this region with the Nelder-Mead algorithm \citep{10.1093/comjnl/7.4.308}. The improvement with respect to the null hypothesis (no ALPs) is quantified by:

\begin{equation}
\Delta \log \mathcal{L} = \log \mathcal{L}_{\rm ALP} - \log \mathcal{L}_{\rm null}.
\end{equation}

\section{Results and Interpretation}

\begin{figure}
    \centering
    \includegraphics[width=\columnwidth]{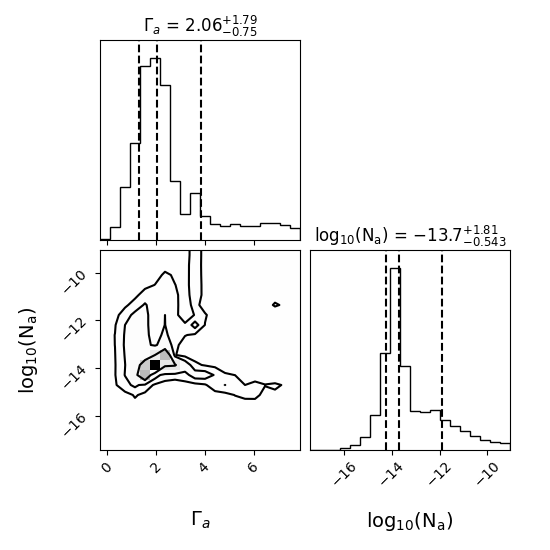}
    \caption{Corner plot displaying the 45,000 independent spectral realizations. It shows the best fits for $\rm{N}_a$ (down-right) and $\Gamma_a$ (up-left). The down-left plot illustrates their correlation distribution, revealing no clear evidence of correlation between the two parameters.}
    \label{fig:cornerplot}
\end{figure}

The fitted spectral parameters of the ALP-photon mixing component obtained for each of the 50 magnetic field realizations can be visualized in Figure \ref{fig:cornerplot}. No clear evidence of a correlation between $\Gamma_{a}$ and $\rm{N}_{a}$ is observed. For most of the realizations, the ALP-photon mixing component results in a spectral index of $\sim2$ and a flux normalization at $3$ TeV of $\sim2\times10^{-14}\;\rm{TeV}^{-1}\;\rm{cm}^{-2}\;\rm{s}^{-1}$. This is an ALPs flux of the order of the photon flux emitted by M87.

\begin{figure}
    \centering
    \includegraphics[width=\columnwidth]{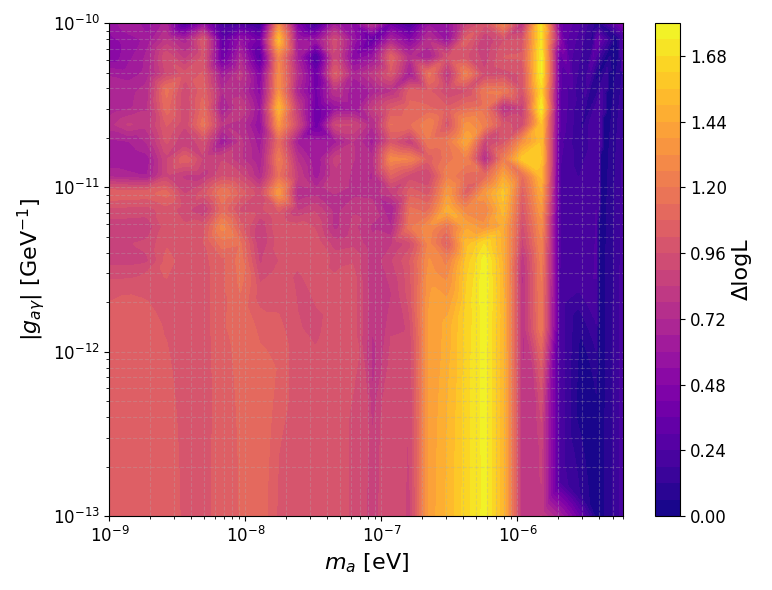}
    \caption{Likelihood improvement $\Delta \log \mathcal{L}$ (as defined in section \ref{subsec:4.3}) between two different models in the ALPs ($m_a$ and $g_{a\gamma}$) parameter space. The color scale represents the maximum likelihood over the null hypothesis (no ALPs) for each candidate. The most favorable region lies around $m_{a}\sim 5\times10^{-7}$ eV and $g_{a\gamma}< 5\times10^{-12}\;\rm{GeV}^{-1}$, corresponding to the yellow region in the parameter space.}
    \label{fig:logL}
\end{figure}

In Figure \ref{fig:logL}, for each candidate \((m_a, g_{a\gamma})\) the highest \( \Delta \log \mathcal{L} \) among the 50 magnetic field realizations is shown. The corresponding spectral parameters \( (N_0, \Gamma, N_a, \Gamma_a) \) are recorded, allowing the exploration of degeneracies between the standard and ALP-induced components.

Although the best candidate for ALP remains within a region excluded by existing experimental constraints, a well-defined surrounding area shows comparable increases in $\Delta \log \mathcal{L}$ to the best candidate (see Fig. \ref{fig:logL}). This indicates that the ALPs parameter space remains viable, with a slight preference for masses around $5.2\times 10^{-7}$ and coupling factors $\leq 5\times 10^{-12}$. These motivate further exploration with a spectrum more precisely measured above tens of TeV.

\begin{figure*}
    \centering
    \includegraphics[width=\columnwidth]{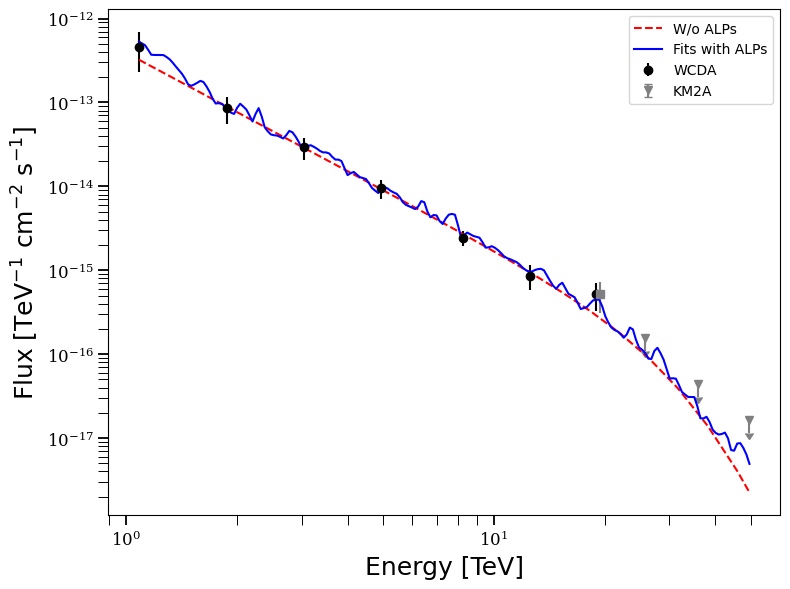}
    \includegraphics[width=\columnwidth]{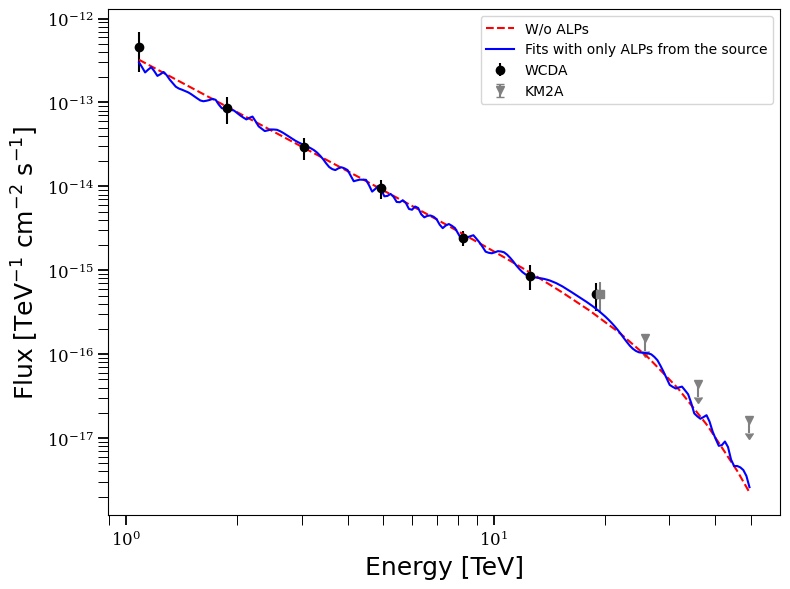}
    \caption{Flux points detected by LHAASO WCDA (black) and KM2A (grey) are shown. The red line represents the best fit to the source spectrum without the ALP-photon mixing contribution. The blue line in the right panel is the resulting spectrum considering only photon-ALP-photon oscillation, while the left panel shows the resulting spectrum considering an extra ALP population.}
    \label{fig:spectrum}
\end{figure*}

The most favorable non-excluded candidate for ALPs (\( m_a = 5.73 \times 10^{-7}~\mathrm{eV} \) and \( g_{a\gamma} = 2.8 \times 10^{-12}~\mathrm{GeV}^{-1} \)) yields a maximum improvement against the null hypothesis of \( \Delta \log \mathcal{L} = 1.78 \), corresponding to a significance of {1.89$\sigma$}. To also compare whether the ALP-photon mixing component is favored we also compute the log-likelihood for a source-only photon-ALP mixing scenario; obtaining for the best non-excluded candidate (\( m_a = 2.78 \times 10^{-7}~\mathrm{eV} \) and \( g_{a\gamma} = 6.81 \times 10^{-12}~\mathrm{GeV}^{-1} \)) an improvement of \( \Delta \log \mathcal{L} = 1.03 \), equivalent to 1.43$\sigma$. The resulting fits for both of the best candidates mentioned are depicted in Figure \ref{fig:spectrum}. 

These findings indicate that, even though the significance of the fit improves with respect to a no ALP-photon mixing scenario, the current available data do not show a statistically significant improvement. This means that a contribution to M87 VHE gamma-ray spectrum from an ALP-photon mixing component is not preferred over a standard physics scenario. 

Nevertheless, it is worth noting that these results were computed using LHAASO's flux upper limits at tens of TeV, and could potentially lead to an improvement with measured flux points. Thus, we also investigate the hypothetical scenario in which the current upper limits are treated as actual flux detections with estimated uncertainties of $10\%$ of the flux (as may be the case with more data in the future). In this scenario, the improvement becomes statistically significant, with the best candidate for ALP yielding an improvement of \(\Delta \log \mathcal{L} = 4.85\), i.e., a significance of approximately $3.11\sigma$. This would represent an indication of a preference for an ALP-photon mixing component to explain the hardening at $\sim20$ TeV in the M87 spectrum. While this result must be interpreted with caution, as it requires future confirmation of tentative flux excess at tens of TeV, it demonstrates that if those spectral regions currently constrained by upper limits are reported as measured flux points in future observations, an ALP-induced photon component may emerge as a plausible explanation for the observed spectral hardening in M87.

\section{Conclusions and Outlook}

We studied the viability of an ALP-photon mixing component as an explanation for the spectral hardening around $\sim 20$~TeV observed by LHAASO in the VHE spectrum of M87. The total expected photon flux under this ALPs scenario was computed and compared with the observed data to obtain the log-likelihood of the fit. We then compared the log-likelihood of each ALP candidate against that of the null hypothesis to evaluate whether including ALP-photon mixing improves the spectral description, quantifying this improvement through the $\Delta \log \mathcal{L}$. 

Our results indicate that ALPs could affect the spectral shape, particularly at energies $E \gtrsim 10$ TeV, where Standard Model predictions imply strong attenuation. Within the explored region of parameter space, we identified ALP candidates that improve the fit to the VHE spectrum of M87. The best result for a non-excluded ALP candidate yields an improvement of 1.89$\sigma$ over the null hypothesis, indicating that it is not statistically significant with the currently available data. Additional observations above energies of 20 TeV could lead to a statistically significant improvement, potentially reaching a significance level exceeding $3\sigma$.

M87 is a promising source for testing physics beyond the Standard Model, such as scenarios involving axion-like particles. The spectral hardening observed by LHAASO, and also by HAWC, may indicate the presence of an additional component that is not easily accounted for by standard emission models. Although an ALP-photon mixing component does not explain this feature with statistical significance, the scenario cannot be ruled out and warrants further observations to assess its viability. When combined with other astrophysical probes, these efforts may contribute to the detection or constraints of ALP properties.

\section*{Acknowledgments}
We acknowledge support from the projects UNAM-PAPIIT IG101323 and SECIHTI LNC-2023-117.

\bibliographystyle{mnras}
\bibliography{sample7}
\label{lastpage}
\end{document}